\documentclass[twocolumn,preprintnumbers,amsmath,amssymb]{revtex4}

\usepackage{graphicx}% Include figure files
\usepackage{dcolumn}% Align table columns on decimal point
\usepackage{bm}% bold math
\usepackage{amssymb}
\usepackage{amsmath}
\usepackage{enumerate}
\usepackage{wrapfig,epsfig}
\usepackage{graphics}
\usepackage{shadow}
\usepackage[T1]{fontenc}
\usepackage{color}
\usepackage{array}
\usepackage{rotating}

\begin{document}

\title{A microfluidic device based on droplet storage\\ for screening solubility diagrams}
\author{Philippe Laval}
\email{philippe.laval-exterieur@eu.rhodia.com}
\affiliation{LOF, unité mixte Rhodia--CNRS--Bordeaux 1, 178 avenue du Docteur Schweitzer, F--33608  Pessac cedex -- FRANCE}
\author{Nicolas Lisai}
\affiliation{LOF, unité mixte Rhodia--CNRS--Bordeaux 1, 178 avenue du Docteur Schweitzer, F--33608  Pessac cedex -- FRANCE}
\author{Jean-Baptiste Salmon}
\affiliation{LOF, unité mixte Rhodia--CNRS--Bordeaux 1, 178 avenue du Docteur Schweitzer, F--33608  Pessac cedex -- FRANCE}
\author{Mathieu Joanicot}
\affiliation{LOF, unité mixte Rhodia--CNRS--Bordeaux 1, 178 avenue du Docteur Schweitzer, F--33608  Pessac cedex -- FRANCE}
\date{\today}
\begin{abstract} 
This work describes a new microfluidic device developed for rapid screening of solubility diagrams. In several parallel channels, hundreds of nanoliter-volume droplets of a given solution are first stored with a gradual variation in the solute concentration. Then, the application of a temperature gradient along these channels enables us to read directly and quantitatively phase diagrams, concentration vs. temperature. We show, using a solution of adipic acid, that we can measure ten points of the solubility curve in less than 1~hr and with only 250~$\mu$L of solution.
\end{abstract}
\maketitle

%*******************************************************************************************************************
%                                                                                                                                                                                               INTRODUCTION
%*******************************************************************************************************************

\section{Introduction}
Chemistry, biology, and pharmacology, are facing always more complex systems depending on multiple parameters. Therefore their complete investigations take time and require significant amounts of products. In this context, robotic fluidic workstations have already met a great success and proved their efficiency for instance in the genome sequencing and analysis \cite{SANDERS2000}. However, these instruments remain very expensive, need important labor, and the volumes involved ($\leq$~mL) are still too large for some specific applications ({\it e.g.} proteomics) \cite{LUFT2001,CHEN2006}.

Nowadays, other high throughput techniques based on microfluidics \cite{STONE2004,SQUIRES2005} can offer suitable alternative solutions for the development of rapid screening tools. Microfluidic devices are now largely used in biological and chemical fields for multiple applications \cite{VILKNER2004} like molecular separations and cells sorting \cite{MINC2004}, polymerase chain reaction \cite{KHANDURINA2002E,CHABERT2006}, rapid micromixing and analysis of chemical reactions \cite{STROOCK2002,KHAN2004,CHAN2005,SALMON2005}\dots\,Moreover, the development of microvalves and micromixers has made possible the production of highly integrated systems which can be used to address individually hundreds of reaction chambers \cite{THORSEN2002}. These devices are well adapted to carry out high throughput screening of phase diagrams, particularly in the case of protein crystallization investigation. However, their fabrication and multiplexing are still complicated. Another possible strategy is the use of droplets playing the role of nanoliter-sized reaction compartments. These droplets can be produced in specific microfluidic geometries \cite{THORSEN2001}, and their volume and chemical composition can be fixed in a controlled way. In addition, they also allow a rapid mixing of the different compounds, prevent from any hydrodynamic dispersion and cross contamination, and can be stored in microchannels (see Ref.~\cite{SONG2006} and references therein). Such a strategy has already proved to be useful for crystallization studies: {\it e.g.} screening of protein crystallization conditions \cite{ZHENG2003,SHIM2006}, or crystal nucleation kinetics measurements \cite{LAVAL2007}. 

\begin{figure}[ht!]
\begin{center}
\scalebox{1.0}{\includegraphics{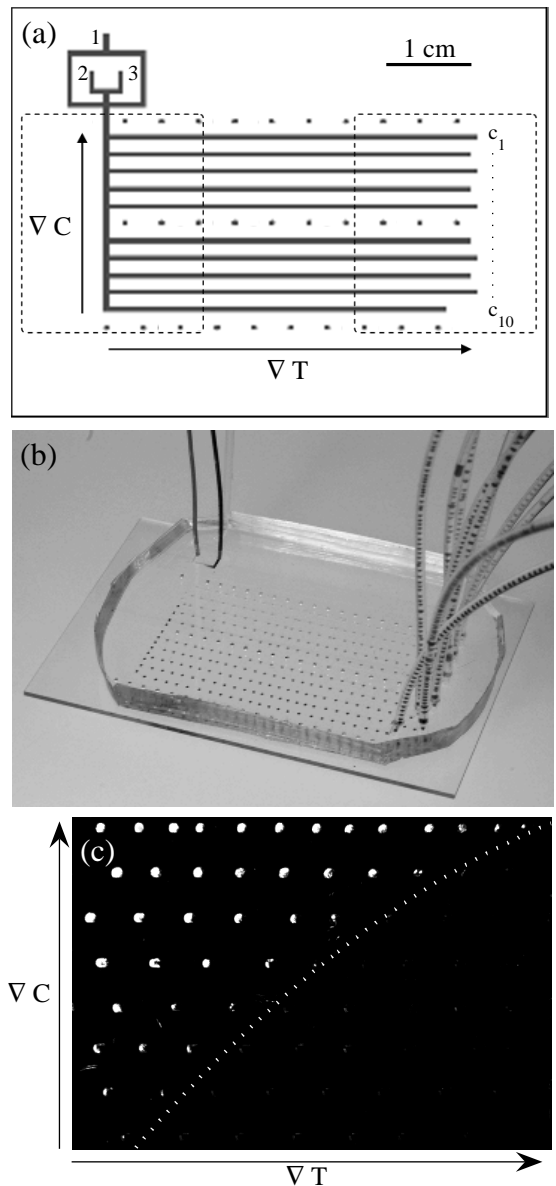}}
\caption{\small (a) Design of the microfluidic device (channels width 500~$\mu$m). Silicone oil is injected in inlet 1 and aqueous solutions in inlets 2 and 3. The two dotted areas indicate the positions of the two Peltier modules used to apply temperature gradients $\nabla T$. The three lines of dots mark the positions of temperature measurements. (b) Picture of the microfluidic chip made of PDMS sealed with a glass slide (76$\times$52~mm$^{2}$) to improve clarity. Droplets containing a colored dye at different concentrations are stored in the ten parallel channels. (c) Example of direct reading of a solubility diagram. The droplets contain an organic solute. The dotted line bounding droplets containing crystals give an estimation of the solubility limit (see section Results for details).}
\label{Puce}
\end{center}
\end{figure}

Figure~\ref{Puce} summarizes the main insights of our work. We have engineered a new microfluidic chip that allow a direct and quantitative reading of two-dimensional diagrams. Hundreds of nanoliter-sized droplets of different chemical compositions can be stored in parallel microchannels,
and a temperature gradient applied along these channels enables us to obtain a two-dimensional array of droplets of different concentrations and temperatures. For solubility diagram screening, droplets containing a given solute are first stored with a gradual variation of concentration. Then, crystallization in the droplets is induced by cooling, and finally, the application of an adequat temperature gradient dissolves crystals in droplets whose temperature is higher than their solubility temperature. As a result, we directly read the limit between droplets with and without crystals as shown on Fig.~\ref{Puce}(c), which gives the solubility temperatures of the solution at the different concentrations.

%As a result, in each storage channel, the temperature of the limit between droplets with and without crystals read on Fig.~\ref{Puce}(c), gives the solubility temperature of the solution at the corresponding concentration.

In the materials and methods section, we describe the microfluidic device, the method used to store the droplets in the channels, and the temperature control setup. We also characterize the concentration and temperature gradients. In the last section, we present an experimental protocol to measure quantitatively solubility diagrams using this device. We demonstrate its efficiency by measuring with only 250~$\mu$L of solution, the solubility curve of an organic compound.

%*******************************************************************************************************************
%                                                                                                                                                                                               MATERIALS AND METHODS
%*******************************************************************************************************************

\section{Materials and Methods}
\subsection{Microfabrication}
The microfluidic device is fabricated in poly(dimethylsiloxane) (PDMS) by using soft-lithographic techniques \cite{MCDONALD2002}. PDMS (Silicone Elastomer Base, Sylgard 184; Dow Corning) is molded on master fabricated on a silicon wafer (3-Inch-Si-Wafer; Siegert Consulting e.k.) using a negative photoresist (SU-8 2100; MicroChem). To make molds of 500~$\mu$m height, we spin successively two 250~$\mu$m thick SU-8 layers on the wafer. After each spincoating process, the wafer is soft-baked (10~min/65$^{\circ}$C and 60~min/95$^{\circ}$C). Photolithography is used to define negative images of the microchannels. Eventually, the wafer is hard-baked (25~min/95$^{\circ}$C) and developed (SU-8 Developer; MicroChem).
A mixture 10:1 of PDMS is molded on the SU-8 master described above (65$^{\circ}$C/60~min). The crossed linked PDMS layer is then peeled off the mold and holes for the inlets and outlets (1/32 and 1/16~in. o.d.) are punched into the material. Then, the PDMS surface and a clean silicon wafer surface (3-Inch-Si-Wafer; 500~$\mu$m; Siegert Consulting e.k.) are activated for 2~min in a UV ozone apparatus (UVO Cleaner, Model 144AX; Jelight) and brought together. Finally, the device is placed at 65$^{\circ}$C for 2~hr to improve the sealing.

\subsection{Droplet storage protocol}
The device, presented on Fig.~\ref{Puce}(a), is composed of three inlets and ten outlets located at the extremities of channels c$_1$ to c$_{10}$. As shown on Fig.~\ref{Puce}(b), each outlet is connected to a $\approx$\,20~cm long rigid tubing (FEP 1/16~in.) ended with a piece of soft PVC tubing (Nalgene, $\approx$\,5~cm long) inserted in an automated pinch electrovalve (105S--01059P; Asco Joucomatic). Thanks to this system, each outlet can be independently closed or opened by pinching or not the corresponding PVC tubing. However, the pinching out of a tube leads in a liquid displacement. To minimize the subsequent liquid disturbance in the microchannels, the electrovalves are placed close to the rigid ones, and the hydrodynamic resistance after the electrovalves is kept as weak as possible using large tubing. 

Silicone oil (500~cSt; Rhodorsil) is injected in inlet 1 at constant flow rate $Q_{1}\approx3$~mL~hr$^{-1}$, and aqueous phases are injected at flow rates $Q_{2}$ and $Q_{3}$ ranging from 0 to about 1~mL~hr$^{-1}$, in inlets 2 and 3 respectively. All liquids are injected with syringe pumps (PHD 2000 infusion; Harvard Apparatus). At the intersection between the oil and the aqueous streams, monodisperse droplets of the aqueous phase in oil are continuously produced \cite{ANNA2003}. Both the droplet volume (about 100--300~nL) and their production frequency (typically between one to ten droplets per second) can be tuned by the ratio of oil to aqueous phase flow rates. The droplet composition is monitored by the ratio $Q_{2}/Q_{3}$. 

Thanks to the possible opening and closing of each outlet, we can store droplets of given aqueous compositions in the different storage channels c$_i$. Several steps are necessary to perform such a filling. First, all the channels are initially filled with silicone oil. Secondly, the outlet of channel c$_1$ is opened and all the others are closed. In this configuration, all the droplets of a given composition flow through c$_1$. Finally, once the flow is stable, the outlet of c$_1$ is suddenly closed and simultaneously, the outlet of channel c$_{2}$ is opened. All the droplets previously present in c$_1$ stay immobilized whereas the other droplets, whose composition can be changed, flow through c$_{2}$. Successively, in the same way, we can store droplets of various compositions in all channels c$_i$. 

\subsection{Chemical composition control}
For solubility investigations, the  control of the concentrations in the droplets is crucial. However, because of PDMS elasticity and syringe pumps precision, an inaccuracy in droplet concentration remains. To estimate this error, we have performed investigations with a confocal Raman microscope (HR800 Horiba; Jobin-Yvon). A $50\times$ microscope objective was used for focusing a 532~nm wavelength laser beam in the droplets, and for collecting Raman scattered light, subsequently dispersed with a grating of 600 lines per millimeter. To minimize the out-of-focus background signals, we fixed the confocal pinhole at 500~$\mu$m. Experiments were performed on droplets made of two initial aqueous solutions of K$_{4}$Fe(CN)$_{6}$ (0.5~M) and K$_{3}$Fe(CN)$_{6}$ (0.5~M) injected in inlets 2 and 3 respectively. These two compounds display strong and distinct Raman signals \cite{CRISTOBAL2006}. 

\begin{figure}[htbp]
\begin{center}
\scalebox{1.0}{\includegraphics{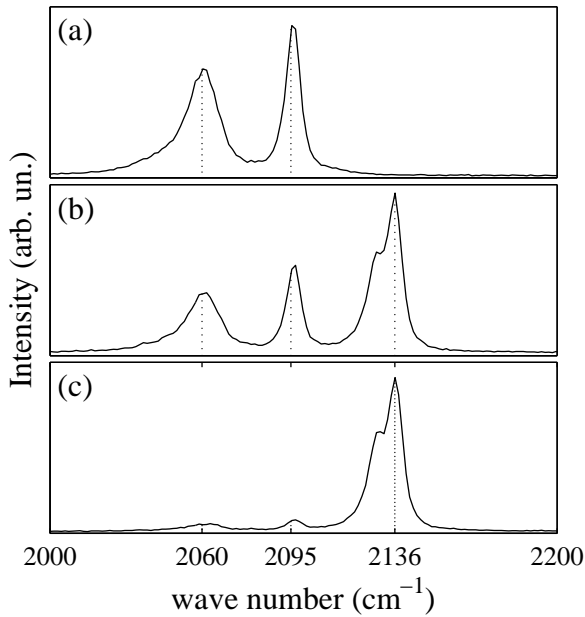}}
\caption{\small Raman spectra of droplets containing different concentration ratios $R_{C}$ of potassium ferrocyanide K$_{4}$Fe(CN)$_{6}$ and potassium ferricyanide K$_{3}$Fe(CN)$_{6}$. (a) $R_{C}=0$ (b) $R_{C}=1$ (c) $R_{C}=9$.}
\label{Spectres}
\end{center}
\end{figure}

Figure \ref{Spectres} shows three Raman spectra measured in droplets containing different concentration ratios 
R$_C$=[K$_4$Fe(CN)$_6$]/[K$_3$Fe(CN)$_6$].
%R$_C$=[K$_4$Fe(CN)$_{6}]/[\texttt{K}_{3}\texttt{Fe(CN)}_{6}]$.
 The two bands centered at 2060 and 2095~cm$^{-1}$ correspond to K$_{3}$Fe(CN)$_{6}$ and the one at 2136~cm$^{-1}$ corresponds to K$_{4}$Fe(CN)$_{6}$. The concentration of each compound can be probed from the area under their specific Raman bands by:
\begin{equation}
A_{i}=K_{i}C_{i}tV\,,
\end{equation}
where $A_{i}$ is the area under the Raman band of the compound $i$, $C_{i}$ its concentration, $K_{i}$ a specific constant, $t$ the acquisition time, and $V$ the analysis volume. As a consequence, the ratio $R_{A}$ of the Raman bands areas of K$_{4}$Fe(CN)$_{6}$ and K$_{3}$Fe(CN)$_{6}$ is proportional to the concentrations ratio $R_{C}$, and does not depend on the acquisition parameters.

\begin{figure}[htbp]
\begin{center}
\scalebox{1.0}{\includegraphics{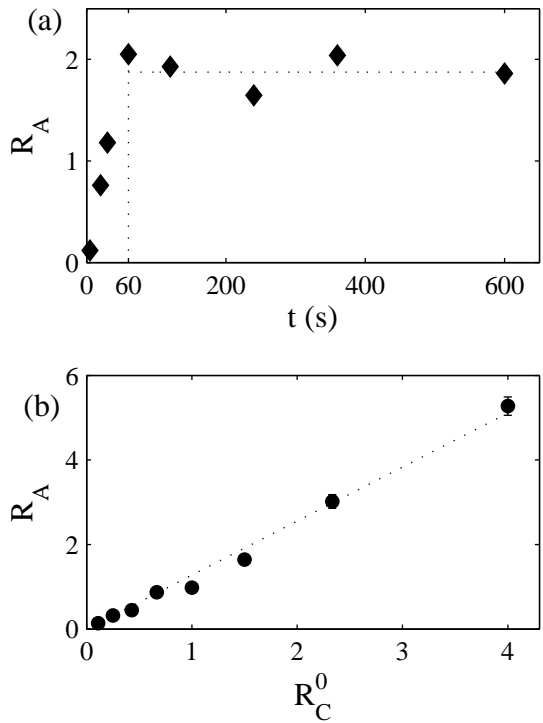}}
\caption{\small (a) Evolution of the ratio $R_{A}$ in the droplets after a sudden change of the aqueous solutions flow rates $Q_{2}$ and $Q_{3}$. Before $t=0$~s, $Q_{2}=0$ and $Q_{3}=500~\mu$L~hr$^{-1}$. For $t>0$~s, $Q_{2}=Q_{3}=250~\mu$L~hr$^{-1}$. Between $5$ and $30$~s, $R_{A}$ are obtained from three single droplets in channel c$_1$. After $t=60$~s, each point is a mean value calculated on several droplets in a given channel. (b) Concentration ratio $R_{A}$ in droplets as a function of the concentration ratio $R^{0}_{C}$ determined from the aqueous solutions flow rates. The dotted line corresponds to the linear fit of the data.}
\label{Concentration}
\end{center}
\end{figure}

In order to optimize the filling protocol, we first use Raman microscopy to follow the kinetics of the concentration stabilization in the droplets after a sudden change in the aqueous phases flow rates. Indeed, due to the PDMS elasticity and the injection system (syringe pumps), the finite response time of the device does not allow instantaneous change of the concentrations. To estimate this response time, we have performed the following experiment: for $t<0$~s, $Q_{2}=0$ and $Q_{3}=500~\mu$L~hr$^{-1}$, and for $t>0$~s, $Q_{2}=Q_{3}=250~\mu$L~hr$^{-1}$. Droplets first flow through channel c$_1$ which is closed after 30~s. Then, droplets are stored in five other channels after 1, 2, 4, 6, and 10~min. Thus, Raman spectra obtained from the droplets in the different channels enable us to follow the evolution of $R_{A}$ as a function of time after the flow rates change. Figure~\ref{Concentration}(a) shows it reaches almost a constant value after 60~s meaning the concentrations become stable after this time. 
Such measurements illustrate that 20~min long protocols are efficient to store 
droplets of desired compositions in the ten channels ($\approx 2$~min per channel).
%the droplets are stored in channel c$_1$ then, every minutes, and up to 9~min, droplets are stored in a new channel. Thus, Raman spectra obtained from the droplets in the different channels enable us to follow the evolution of $R_{A}$ as a function of time after the flow rates change. Figure~\ref{Concentration}(a) shows it reaches almost a constant value after 60~s meaning the concentrations become stable after this time. 

A second series of experiments was performed to estimate and characterize the concentration gradient we can apply in the device. The storage channels are filled  with droplets of different concentrations in K$_{3}$Fe(CN)$_{6}$ and K$_{4}$Fe(CN)$_{6}$ set from the flow rates. In each channel c$_i$, to reach a stable droplets composition, we maintain the flow for 90~s before closing the outlet to store them [see Fig.~\ref{Concentration}(a)]. By measuring the Raman spectra of the droplets composition in the different channels, we obtain the ratio $R_{A}$ as a function of the theoretical ratio $R^{0}_{C}=Q_{2}/Q_{3}$. The error bar corresponds to the standard deviation of the measurements performed on the droplets in a given channel. As can be seen on Fig.~\ref{Concentration}(b), a linear relationship between $R_{A}$ and $R^{0}_{C}$ is observed as expected. Deviations of a few percents around the linear law are probably due to the Raman measurements uncertainties, to the accuracy of the injection system, and also to the PDMS elasticity.

These Raman measurements demonstrate that with the developed protocol, we are able to store hundreds 
of droplets in ten channels in about 20~min, and consuming less than a few hundreds of $\mu$L of solution.
We believe that more rigid and smaller microdevices combined with even more reactive injection system 
would decrease signicantly the amount of liquids used when filling the channels. Other strategies involving for instance
droplet generation thanks to integrated microvalves \cite{LAU2007}, may also proved to be useful 
to decrease the required volumes of solution.

\subsection{Temperature control}
\begin{figure}[htbp]
\begin{center}
\scalebox{1.0}{\includegraphics{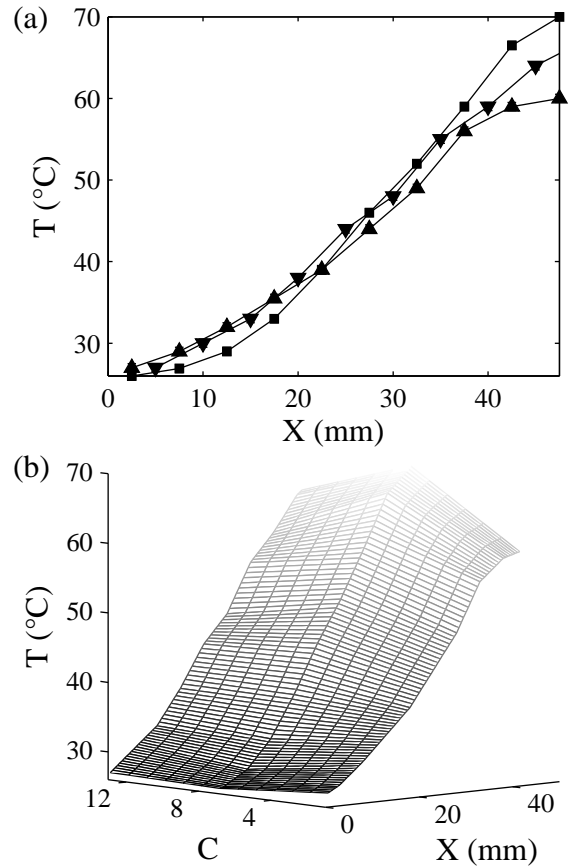}}
\caption{\small Temperature profiles of the chip for a given temperature gradient (a) Temperatures measured along the storage channels with thermocouples inserted through the PDMS layer at different positions shown on Fig.~\ref{Puce}(a). ($\blacktriangle$) measurements series above channel c$_1$; ($\blacksquare$) series between c$_{5}$ and c$_{6}$; ($\blacktriangledown$) series below c$_{10}$. (b) Interpolated temperature profile of the chip.}
\label{Temperature}
\end{center}
\end{figure}

The temperature field of the chip is controlled with two Peltier modules (30$\times$30$\times$3.3~mm$^{3}$; CP1.4--71--06L; Melcor) placed underneath the wafer at positions marked by the two dotted areas on Fig.~\ref{Puce}(a). Since the two Peltier modules are independant, we can heat or cool the device, and also apply important temperature gradients. We use a silicon wafer as chip support to optimize thermal transfers and thus to create regular temperature gradients along the storage channels. Thin thermocouples (type K, 76~$\mu$m o.d., 5SRTC-TTKI-40-1M; Omega) measure the temperature of the device along three series of positions parallel to the storage channels. The first series is placed above c$_1$, the second one between c$_{5}$ and c$_{6}$, and the third one below c$_{10}$ [see Fig.~\ref{Puce}(a)]. To reach the maximal precision on liquid temperature measurements inside the channels, the thermocouples are inserted in holes previously punched through the PDMS layer and filled with silicone oil. Thermocouples signals are processed with a data acquisition instrument (USB--9161; National Instruments) and LabView software. Figure~\ref{Temperature}(a) shows we are able to apply easily temperature gradients up to 45$^\circ$C on 5~cm.
To estimate the temperature at any positions along the storage channels, we perform a longitudinal and transverse linear interpolation of the three series of measurements. The final profile obtained after such interpolation is depicted on Fig.~\ref{Temperature}(b). Note that the temperature is not perfectly homogeneous transversely to the storage channels.
This is due to the size of the Peltier module 
as compared to the size of the droplet storage area: smaller storage area, or larger Peltier modules,
would give homogeneous temperature profiles along the transverse direction of the channels.

%*******************************************************************************************************************
%                                                                                                                                                                                                               RESULTS
%*******************************************************************************************************************

\section{Results}
In the previous section we have shown that our microdevice allows us to build a two-dimensional array of droplets with both concentration and temperature gradients. We now present an application of this chip by measuring the solubility curve of an organic solute.

Such measurements are carried out with an adipic acid solution previously prepared in a beaker. It is made of 10.14~g of adipic acid (99\%; Aldrich) in 50.66~g of deionized water. The solubility temperature of the solution is 63$^{\circ}$C. To avoid any crystallization before the droplets formation, the syringe containing the solution and the corresponding tubing are heated at about 65$^{\circ}$C with two flexible heaters (Minco) controlled with temperature controllers (Minco). A stereo microscope (SZX12; Olympus) with an objective (DF PLFL $0.5\times$ PF; Olympus) enables us to observe the device during the solubility study.

%To observe and measure the solubility curve with the microfluidic device, droplets are first stored in the different channels c$_i$ with a gradual variation of concentration in the investigated solute. Then, crystallization in the droplets is induced by cooling, and finally, the application of an adequat temperature gradient along the storage channels dissolves crystals in droplets whose temperature is higher than their solubility temperature. As a result, in each storage channel c$_i$, the temperature between droplets with and without crystals gives the solubility temperature of the solution at the corresponding concentration.

We inject the adipic acid solution in inlet 2 and deionized water in inlet 3. By changing the flow rates ratio we fill the storage channels with droplets whose concentration in adipic acid varies from 20~g~/~100~g of water in channel c$_1$ down to 6~g~/~100~g of water in c$_{10}$. The massic concentration $C$ in the droplets is calculated according to:
\begin{equation}
C = \frac{C^{0}}{1+(1+C^{0})Q_{3}/Q_{2}}
\end{equation}
where $C^{0}$ is the massic concentration of the initial adipic acid solution, $Q_{2}$ and $Q_{3}$ the 
respective flow rates of the solution and water (we checked that density variations induced by the presence of adipic
acid are negligible). The microfluidic chip is kept at about 65$^{\circ}$C using the Peltier modules to avoid any crystallization during the droplet storage. Before stopping the droplets in a channel, we maintain it open for 90~s for flow stabilization. In these conditions, the total filling of the ten channels is reached in less than 20~min and only 250~$\mu$L of solution are spent.

After the droplet storage, crystallization is induced by cooling. Note that the mean time of 
crystal nucleation is inversely proportional to the reactor volume. 
Indeed, the nucleation frequency is given by $1/JV$ where $J$ is the nucleation rate that does no depend on the volume
$V$ of the reactor (see Refs.~\cite{ZETTLEMOYER1969,KASHCHIEV2003A,MULLIN2001,LAVAL2007} and references therein).
Crystal nucleation in a droplet of 100~nL is thus 10$^{4}$ times longer than in a vial of 1~mL. 
To reduce such long induction time, we apply a strong cooling to increase significantly
the supersaturation. In our case, down to $\approx-5^\circ$C, crystals appear in all the droplets after a few minutes.

\begin{figure}[htbp]
\begin{center}
\scalebox{1.0}{\includegraphics{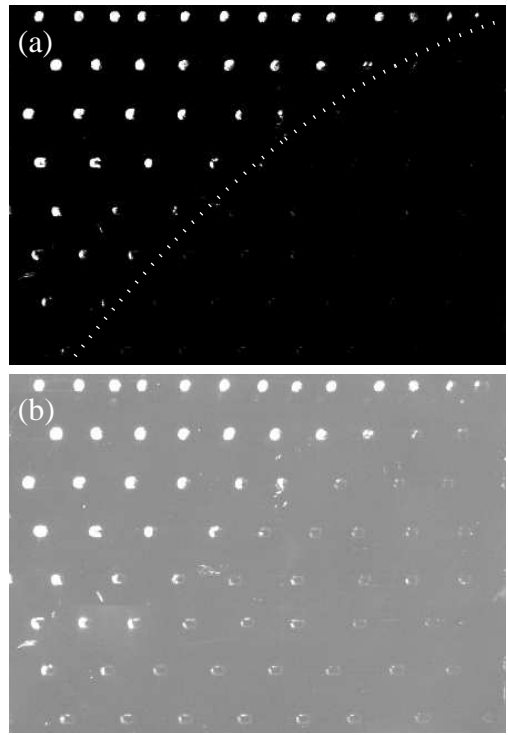}}
\caption{\small Images of a part of the storage area obtained under crossed polarizers. Droplets of adipic acid solution are stored in the channels. The concentration in adipic acid was gradually changed between the upper and the bottom channels. After crystallisation of all the droplets, a temperature gradient is applied (low temperature on the left and high temperature on the right). (a) The dotted line separating droplets containing crystals from empty droplets give an estimation of the solubility limit. (b) Same image but with a different contrast displaying the droplets positions.}
\label{Gouttes}
\end{center}
\end{figure}

\begin{figure}[htbp]
\begin{center}
\scalebox{1.0}{\includegraphics{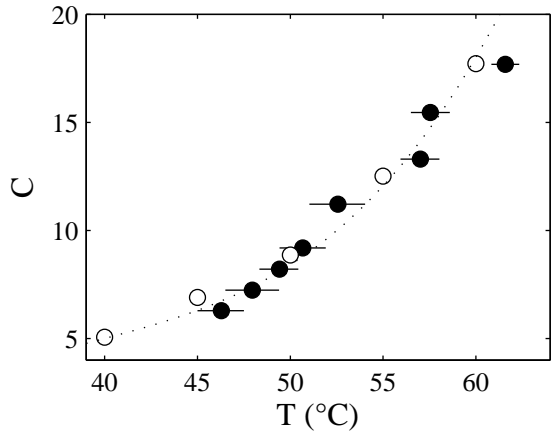}}
\caption{\small ($\bullet$) Solubility of adipic acid in water measured in the case of a temperature gradient of 0.7$^{\circ}$C~mm$^{-1}$. ($\circ$) Solubility data from literature, the dotted line is a guideline for eyes.}
\label{Solubility}
\end{center}
\end{figure}

To obtain the solubility curve directly on the chip, we then apply a temperature gradient between 32 and 65$^{\circ}$C after the crystallization step. Crystals dissolve in all the droplets whose temperature is higher than their solubility temperature. In the other droplets, crystals are partly solubilized but still exist (the equilibrium
is reached in about 20~min). Typical images of the storage area are presented on Fig.~\ref{Gouttes}.
Since adipic acid crystals have birefringent properties, they are easily detected under crossed polarizers.
The smallest detectable crystals size is about 50$\times$50~$\mu$m$^2$ at the magnification used. Figure~\ref{Gouttes}(a) enables us to directly observe the limit of crystal presence. 
Using interpolated temperature profiles such as the one displayed in Fig.~\ref{Temperature}, allows us
to estimate the solubility temperatures for all the ten concentrations (we choose them in the middle of the two sucessive droplets with and without crystals). Figure~\ref{Solubility} presents such solubility temperatures measured with our microfluidic device. The error corresponds to the temperatures difference between the two droplets enclosing the solubility limit positions.
These results are in good agreements with data obtained from literature \cite{APELBLAT1987}.

Naturally, the errors done on such measurements depend on the distance between two successive droplets, and on the amplitude of the temperature gradient. In our case, the temperature gradient of 0.7$^{\circ}$C~mm$^{-1}$ and a typical distance of $3$~mm between two droplets give an error of $\pm1^{\circ}$C. The application of smaller temperature gradients and the reduction of the distance between two successive droplets would give a better accuracy on the solubility limit.

For the moment, the maximal temperature which can be investigated is limited by the evaporation of water through the PDMS layer \cite{LENG2006}. Simple measurements show that
the volume of an aqueous droplet stored in our device at 60$^{\circ}$C, decreases by $\approx 10$\% in  4~hr. 
Such an effect is negligible for the experiments described above (droplet filling time 20~min at 65$^{\circ}$C), but may explain the small discrepancy observed on Fig.~\ref{Solubility} at high temperature. We believe that the use of non-permeable materials such as glass, instead of PDMS, could easily broaden the possibilities offered by our system.

%*******************************************************************************************************************
%                                                                                                                                                                                                               CONCLUSION
%*******************************************************************************************************************

\section{Conclusion}
In this work we have presented a new microfluidic tool to perform rapid screening of solubility diagrams. The device enables us to store hundreds of droplets ($\approx 100~$nL) of various chemical compositions in parallel microchannels, and to apply large temperature gradients. We have demonstrated using a model system (adipic acid in water), that we could easily and directly access to ten simultaneous measurements of the solubility curve on a large temperature range, in less than 1~hr, and with only 250$~\mu$L of solution. To conclude, we believe our device is a suitable tool for solubility diagrams screening, more rapid, with a better temperature control, and cheaper than classical robotic workstations. Such a microfluidic tool may also be useful for many other applications, where two-dimensional 
screening, temperature vs. composition, is required.

\acknowledgments{We gratefully thank G.~Cristobal, J.~Krishnamurti, J.~Leng, and F.~Sarrazin
for fruitful discussions and critical reading of this manuscript. 
We also acknowledge {\it Région Aquitaine} for funding and support, and the {\it Atelier Mécanique} of the CRPP for their technical help.}

%\bibliography{N:/users/plaval/PUBLIC/Elements_envoyes/biblioArticle}

\end{document}